\documentclass[a4paper]{jpconf}

\usepackage{graphicx}
\usepackage{epstopdf}
\usepackage{iopams}

\newcommand{\mk}{\langle k \rangle}
\newcommand{\ep}{\varepsilon}

\begin{document}
\title{Living on the edge of chaos: minimally nonlinear models of genetic regulatory dynamics}

\author{Rudolf Hanel$^1$, Manfred P\"ochacker$^1$ and Stefan Thurner$^{1,2}$}

\address{$^{1}$Section for Science of Complex Systems, Medical University of 
Vienna, Spitalgasse 23, A-1090 Vienna}
\address{$^{2}$Santa Fe Institute, 1399 Hyde Park Road, Santa Fe, NM 87501, USA }

\ead{thurner@univie.ac.at}

\begin{abstract}
Linearized catalytic reaction equations -- modeling e.g. the dynamics of 
genetic regulatory networks -- under the constraint that expression levels, i.e.  
molecular concentrations of n€€ucleic material are positive, exhibit nontrivial dynamical properties,
which depend on the average connectivity of the reaction network.  
In these systems the inflation of the {\it edge of chaos} and multi-stability have been demonstrated to exist. 
The positivity constraint introduces a nonlinearity which makes chaotic dynamics possible. Despite the simplicity of such {\it minimally nonlinear systems}, 
their basic properties allow to understand fundamental dynamical properties of complex biological reaction networks. 
We analyze the Lyapunov spectrum, determine the probability to
find stationary oscillating solutions, demonstrate the effect of the nonlinearity on
the effective in- and out-degree of the {\it active interaction network} and study how the
frequency distributions of oscillatory modes of such system depend
on the average connectivity.
\end{abstract}




\section{Introduction}

Many complex systems in general -- and living systems and cells in particular -- display
remarkable stability, i.e. a capacity to sustain their spatial and temporal 
molecular organization. Yet, their stability is dynamic, i.e. 
these systems -- to a certain degree -- are capable of adapting to changes in their 
physical and chemical environment.
This has led several authors \cite{langton,kauffman_order,mitchell,packard}
to interpret such systems as existing at the {\it edge of chaos}.
Mathematically the {\it edge of chaos} refers to regions in parameter space, 
where the system dynamics is characterized by a maximal Lyapunov exponent (MLE), 
$\lambda_1$, equal
to zero. 
In this case small changes in parameters may cause
the dynamics to switch between regular and 
chaotic behavior, thereby being able to {\it explore} large portions of the system's phase-space.
This possibility is most relevant for living systems 
existing in fluctuating environments.
In many dynamical systems the edge of chaos exists only for a tiny portion of parameter space, typically in sets of singular points, i.e. sets of measure zero. 
The dynamics of systems at the edge of chaos can become highly nontrivial, even for 
simple maps like the logistic map \cite{robledo}.
It has been argued that living systems have evolved towards the edge of chaos
by natural selection \cite{kauffman_order}, however 
it is not clear which mechanisms allow 
self-organization around these exceptional regions in parameter space.

Living systems have exist in the state of quasi-stationary nonequilibrium and therefore
can not be closed systems. They require a flow of substrate and energy 
to and from the system. 
Since long \cite{lotka}, rate equations for molecular dynamics have been considered.
For systems to be self-sustaining, such rate equations need to be autocatalytic, i.e.
some molecular species directly or indirectly catalyze their own production.
For living systems, cells in particular, to be in a stationary state, 
production, decay and flow rates of intercellular components effectively have to balance each other, \cite{pross,pross2}. 
Replicating, living systems therefore in general balance between
stationary states (nonreplicating modes) and 
growth (replicating mode), limited by constraints posed by the environment. This balance provides a natural selection criterion.

Autocatalytic systems 
are frequently governed by nonlinear equations  for enzyme-kinetics, e.g. Michaelis-Menten differential equations
\cite{MichaelisMenten1913}, or more general replicator equations,
see e.g. \cite{hofbauer}. 
For various reasons linearized autocatalytic networks have been considered, for the case of abundant 
substrate, see e.g. \cite{JainKrishna,JainKrishna2}, or   
for reverse engineering \cite{yeung,SHT_reverse}.
Systems with linearized dynamics can be easily depicted in terms of directed reaction networks, where
nodes represent molecular species. 
Two nodes, where one node directly influences (production or inhibition) the other, are connected by a directed link.  
Weights of such links quantify associated reaction rates; negative rates indicate inhibitory links.
Weights of self-loops in the reaction network, i.e. links of a node onto itself, quantify decay rates. 
Recent progress in genomic and proteomic technology begins to reveal facts about regulatory networks of 
organisms. There is some evidence that these directed networks show scale-free  
topological organization  \cite{sneppen,jeong,jeong2,babu2004}.  More recent evidence suggests 
topological differences between in- and out-degree distributions 
\cite{barabasi_xx, barabasi_x}.
Basically two main approaches for modeling catalytic networks have been pursued:
Discrete state approaches, e.g. Boolean networks \cite{kauffman}, and continuous approaches, relying on 
ordinary or stochastic differential equations \cite{smith,mahaffy,mestl,chen,chen2}. 
The relevance of noise has been experimentally demonstrated 
\cite{ko,fiering,hasty,haitzler,guptasarma}. 

Interestingly, various models of disordered recurrent networks \cite{kauffman,andrecut}
seem to share three distinguished modes of operation: 
(i) stable, (ii) critical, and (iii) chaotic super-critical. 
These properties could be generic or even {\em universal}. 
The importance of determining the minimum complexity of models exhibiting these 
properties has been pointed out \cite{kauffman} and the question has been raised 
whether these properties can already be found in linear systems.
Following this philosophy we have recently introduced a model for genetic regulatory dynamics \cite{SHT_inflation}.
This model is governed by sets of linear equations 
\begin{equation}
\frac{d}{dt} x_i = \sum_j A_{ij} x_j + J_i + \nu_i \quad ,  
\label{model_raw}
\end{equation}
where $A_{ij}$ is the weighted adjacency matrix of the {\it full} autocatalytic reaction network, 
whose entries may be zero, positive and negative -- indicating that $i$ either has no influence on $j$
or the production of molecular species $i$ is stimulated or suppressed by $j$, respectively.
This means that if substrate $j$ exists, $i$ gets produced (or reduced) at rate $A_{ij}$.
$x_i$ is the concentration of the molecular species $i$ (e.g. proteins or mRNA). 
$J_i$ corresponds to a flow-vector.
The molecular species $i$ flows into the system if $J_i>0$ and out of the system if $J_i<0$.
$\nu_i$ is a suitable noise term.
Negative molecular concentration values $x_i$ do not make any sense, hence we impose
the {\it positivity condition} 
\begin{equation}\label{poscond}
x_i\geq 0\ ,\quad \mbox{for all $i$}\ .
\end{equation} 
In particular if $x_i=0$ and Eq. (\ref{model_raw}) gives $\dot x_i\leq 0$, then effectively $\dot x_i=0$.
Therefore, the concentration $x_i$ will remain zero until Eq. (\ref{model_raw}) gives $\dot x_i>0$.
We refer to this as the {\it minimally nonlinear} (MNL) model.

MNL models have nontrivial properties \cite{SHT_inflation}: 
(i) They have a possibility for chaotic dynamics. 
(ii) MNL exhibit an {\it inflated edge of chaos}. The 
positivity condition causes the small neighborhood of a singular point in parameter 
space (linear system without positivity condition), with MLE $\lambda_1\sim 0$, to form an extended region (plateau). 
This effect gives random strategies of evolutionary phase-space sampling
a finite chance of locating this particular region in parameter space. 
This may offer an explanation for why and how complex chemical 
reaction systems may have found the vicinity of the edge of chaos at all, 
before evolutionary self-organization could take over for an eventual fine tuning.  
(iii) MNL models show {\it multi-stability}.
Perturbations (or moderate noise levels) can push the system from one attractor of the dynamics to another.

These facts raise interesting questions. (a) The existence of chaotic dynamics
in MNL systems straight forwardly suggests to analyze the Lyapunov spectrum of the dynamics, which
encodes information about the attractor of the dynamics.
Numerical simulations indicated so far that MNL systems exhibit several properties, which are particularly
interesting for modeling living systems. (b) 
Can topological differences between in- and out-degree distributions 
\cite{barabasi_xx, barabasi_x}, 
be explained by MNL dynamics? 
MNL dynamics can down-regulate the concentration $x_i$ of a 
fraction of nodes $i$ to zero. These nodes $i$ then cease 
to play an active role in the dynamics of the MNL system.
The 
remaining nodes continue to play an active role in the dynamics
and constitute the {\it active} regulatory reaction network. 
The active network may have topological properties 
that differ from the full network. 
(c) How probable is it to find oscillating dynamics in MNL systems, and how are fundamental frequencies 
of oscillatory dynamics distributed? 
MNL dynamics frequently shows oscillatory dynamics. 
This is particularly interesting, since
periodic dynamics are well known in regulatory networks in the context of
the cell-cycle, e.g. \cite{cellcycle}, or circadien clocks \cite{circadian}.
Evidence has been presented that oscillating regulatory networks are also 
involved in the morphogenesis of mice \cite{ontogenesis}. Moreover, 
eukaryotic cells may encode information about extracellular environment in the frequency 
of stochastic intracellular events, rather than in the concentrations of molecular
species \cite{periodic_xxx}. Intracellular dynamics in terms of (stochastic) rhythmic burst, 
may be a common mechanism of intracellular information transduction .

In Section \ref{summary} we give a summary of the model \cite{SHT_inflation}. 
In Section \ref{results} we report results on the properties of MNL systems.
In Section \ref{conclusions} we conclude.

\section{The stochastic MNL model}\label{summary}

We present the MNL model as introduced in \cite{SHT_inflation}.
There we derived Eq. (\ref{model_raw}) by linearizing a
set of nonlinear differential equations 
\begin{equation}\label{nleq}
\frac{d}{dt} y_i=F_i(y)\quad, 
\end{equation}
where the state vector $y$ represents a
collection of concentrations $y_i$ of molecular species $i$. These molecular species 
include both mRNA and proteins. The state vector $y$ can be written as 
$y=(x_1,\dots x_n, p_1,\dots p_m)$, where the $x_i$, with $i=1\dots m$, are concentrations of mRNA and the 
$p_r$, with $r=1\dots m$, are protein concentrations. 
Equation (\ref{nleq}) can be linearized around some fixed point 
$y^0=(x^0_1,\dots x^0_n, p^0_1,\dots p^0_m)$.
The variables $p_r$ can be eliminated by
the assumption that, around a fixed point, changes of mRNA (or protein) concentrations, 
$\delta x=x-x^0$, translate linearly into variations of the protein (or mRNA) concentrations, 
$\delta p=p-p^0$, i.e. $\delta p_r=\sum_i C_{ri} \delta x_i$, for
some fixed matrix $C$. Assuming (thermal) fluctuations of the
production and degradation rates around average values $A_{ij}$, using the law of large numbers, finally leads to the equation
\begin{equation}
\frac{d}{dt} x_i = \sum_j A_{ij}( x_j - x^0_j) + \xi_{i}(t) (x_i-x^0_i) + \eta_i \quad ,   
\label{model}
\end{equation}
where $\xi_i\in N(0,\bar \sigma)$ and $\eta_i\in N(0, \sigma)$ are independent normally distributed zero-mean 
random variables with
standard deviations $\bar \sigma$ for the multiplicative noise and $\sigma$ for the additive noise.
Comparing with Eq. (\ref{model_raw}) the 
noise-term $\nu_i$ can be identified, $\nu_i=\xi_{i}(t) (x_i-x^0_i) + \eta_i$,
and the flow-vector $J$ is 
\begin{equation}\label{current}
J_i=-\sum_j A_{ij}x^0_j\quad.
\end{equation}
Time series of mRNA expression levels $x_i(t)$ typically oscillate around 
fixed points (average values) $x^0=\langle x_i(t)\rangle_t$.
This can be used to directly feed characteristic mRNA expression profiles into the MNL model.
Although from a purely mathematical point of view, the fixed point $x^0=0$ is a perfectly 
legitimate choice, it contradicts the fact that living systems are open systems and require 
non-vanishing effective flow-vectors, $J\neq0$. Equation (\ref{current}) immediately implies
that the choice $x^0=0$ is incompatible with this requirement.
Here we use $x^0_i=1000$, for all $i$. 

The weighted adjacency matrix $A_{ij}$ can be used to incorporate topological information
on biological networks. 
In our model the decay rates, $A_{ii}<0$, have identical value, for all $i$. 
This assumption is wrong in general but reasonable for mRNA encoding groups of proteins that act
together in stoichiometric complexes 
\cite{mRNA_decay}. 
Since $A$ is largely unknown experimentally
we are interested in
random ensembles of matrices $A$, which can be parametrized with only a few parameters.
We model $A$ as a random matrix in the following way. 
Using terminology from network theory, the {\it out-degree} $k_i$ of a node ($\equiv$ molecular species) $i$ is defined as the number of products $j$ that can be regulated 
by the molecular species $i$.
The {\it in-degree} is the number of molecular species that regulate $i$.  
The ensemble of interaction networks can now be  specified by the 
in- and out-{\it degree distribution} $p_{in/out}(k)$.
Although in principle in- and out-degree distribution can be chosen independently, 
we consider identical in- and out-degree distributions. In \cite{SHT_inflation}
we have compared scale-free networks with Erd\"os-R\'enyi networks \cite{er1959} and have noted only minor effects on the stability, i.e. the formation, of the $\lambda_1\sim 0$ plateau.
Here we will only consider Erd\"os-R\'enyi networks. The associated topological ensembles 
for the full reaction networks are completely specified by the number $N$ of molecular species and the number $L$ of links between them, i.e. the {\it average degree} $\mk= L/N$ of the network.

Once the topology of a network
is fixed, the actual weights $A_{ij}$ are sampled from a normal distribution $N(0,\sigma_A)$
with zero mean and standard deviation $\sigma_A$.
This assumption is experimentally supported by e.g. \cite{dhaeseler}.
We define the constant $D\equiv -A_{ii}/\sigma_A$ and set $\sigma_A=1$ 
in all numerical simulations. The time-increment used for all 
numerical simulations is $dt=0.1$. 

%
%
\begin{figure}[htbp]%
\hspace{2.0cm}
\vspace{0.0cm}
\begin{minipage}[b]{0.6\textwidth}
\includegraphics[width=\columnwidth]{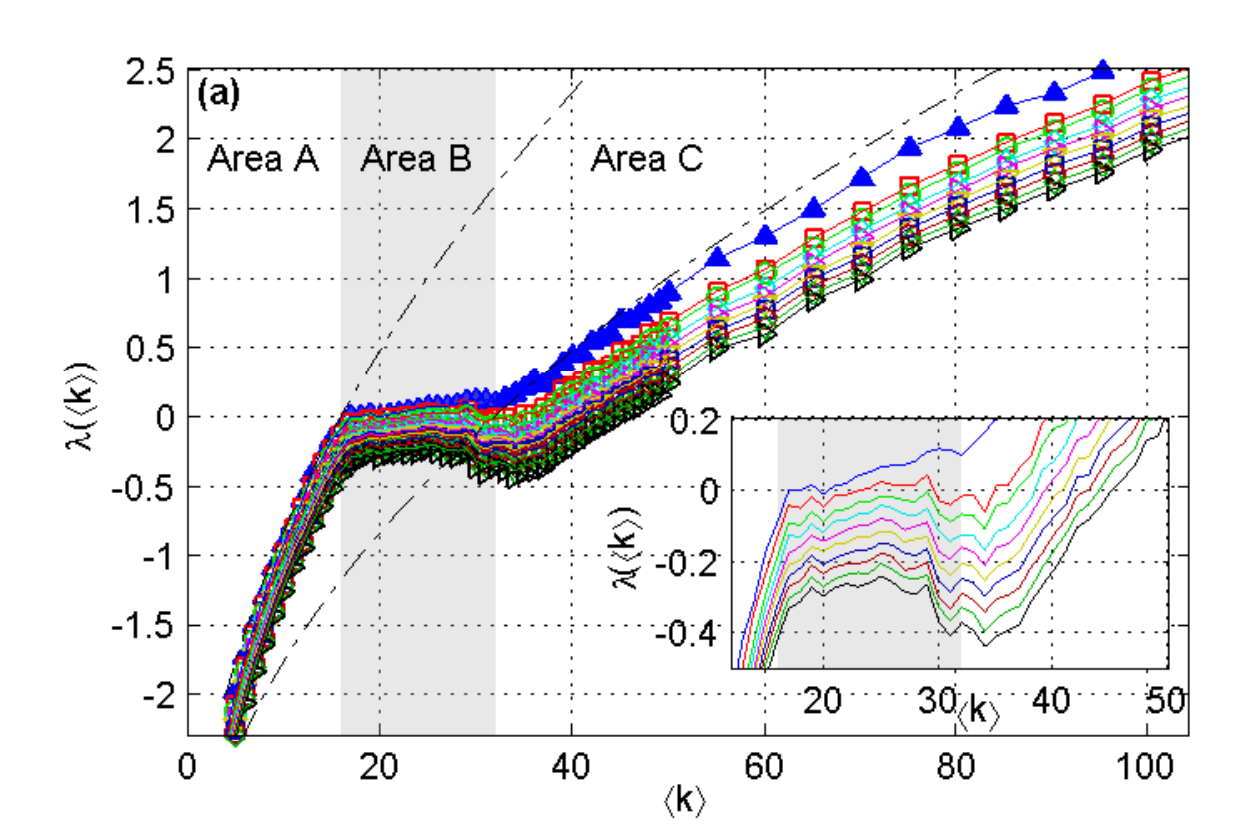}\\ 
\vspace{0.1cm}\hspace{0.59cm}%
\includegraphics[width=0.905\columnwidth]{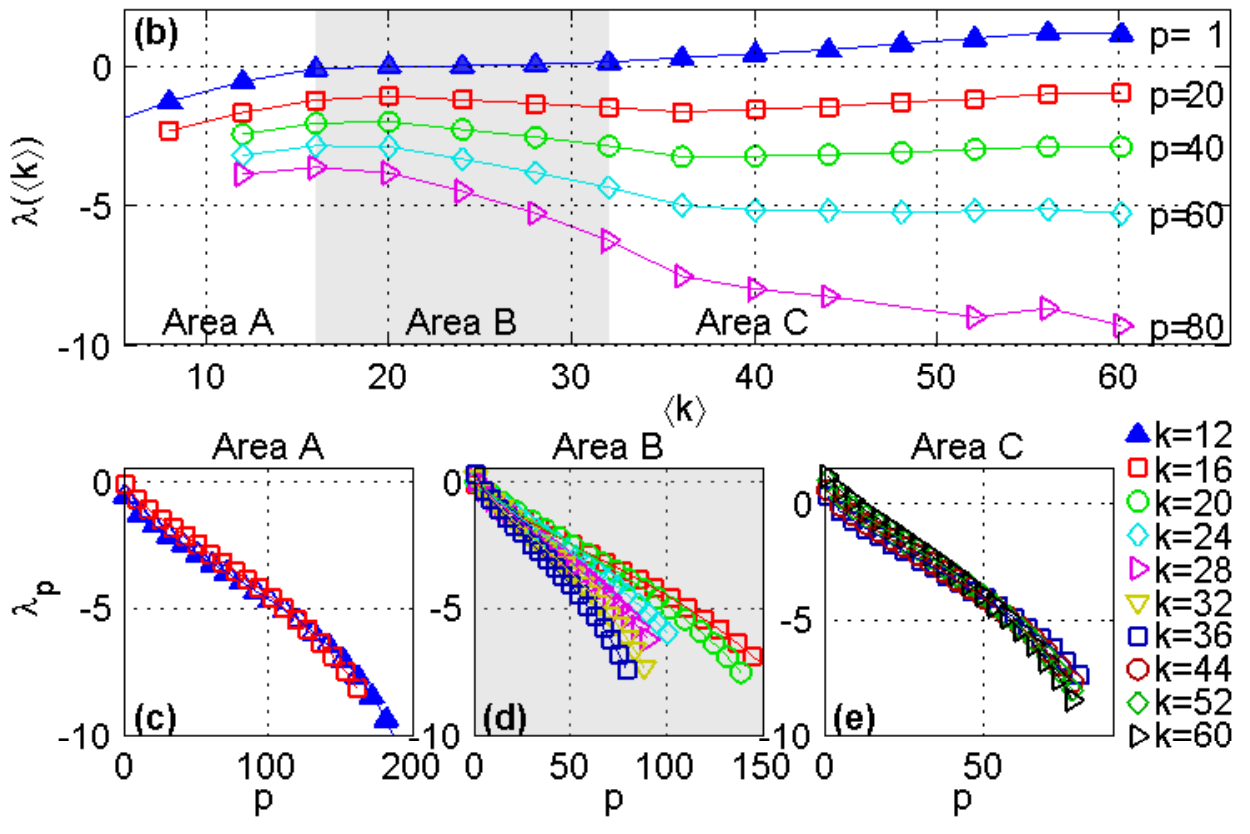} \\  
\vspace{-0.1cm}\hspace{0.34cm}%
\includegraphics[width=0.995\columnwidth]{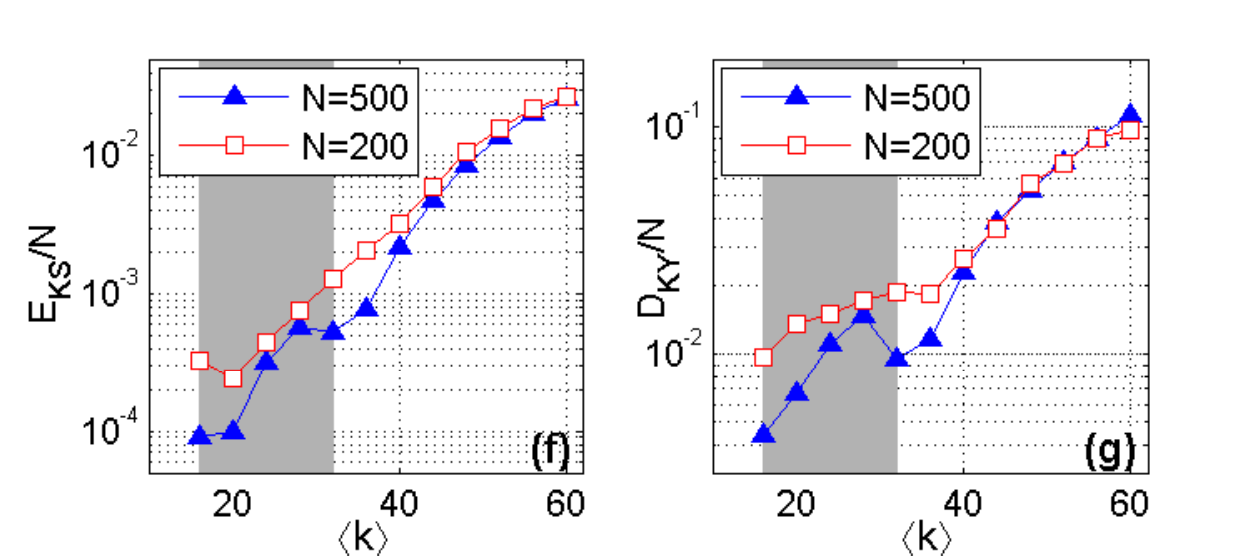} %
\end{minipage} 
	\caption{ 
	(a) Largest ten Lyapunov exponents ($\lambda_p$, $p=1,\dots,10$) of the Lyapunov spectrum ($N=500$).
  The inset magnifies the plateau region. 
	The two black dashed lines are theoretical curves, derived in \cite{SHT_inflation}, 
	approximating $\lambda_1(\mk)$ in the areas A and C.    
	The intersection of these curves with the x-axis, $\lambda_1(\mk)=0$, estimate the beginning and end of the
	$\lambda_1(\mk)\sim 0$ plateau (area B).
	(b) The Lyapunov spectrum as a function of $\mk$ is shown ($N=200$).
	(c,d,e) Lyapunov spectra, $\lambda_p(\mk)$ as functions of $p$, for 
	areas A, B, and C. 
  In area A spectra are all similar, in area B the slope
	of $\lambda_p$ gets steeper with growing $\mk$, while $\lambda_1(\mk)\sim0$. Comparing 
	(d) and (e) shows a clear qualitative difference between area B and C.
	The Kolmogorov Sinai Entropy $E_{KS}$, diagram (f), and the Kaplan York Dimension $D_{KY}$, diagram (g),
	both divided by system size $N$, in logarithmic scale, for $N=200$ and $N=500$.  
  $D_{KY}$ represents an upper bound for the information dimension of the system.  
  $E_{KS}$ is calculated from Pesin's theorem, as the sum over all positive $\lambda_p>0$ in the Lyapunov spectrum,
  for fixed $\mk$.
  \\
  \hfill
	In (a-g) averages are taken over 50 random realizations, time interval $[200,\, 1000]$,
	noise $\sigma=\bar{\sigma}=0.1$, $\sigma_A=1$, $D=4$, and $x^0=1000$.
 	}%
	\label{fig:LyapSpectrum}
\end{figure}
%
%

The maximal Lyapunov exponent, $\lambda_1$, measures the exponential rate with which a perturbation 
of a trajectory propagates over time. If 
$\lambda_1<0$, the perturbation vanishes. If $\lambda_1>0$, the perturbation grows exponentially.
In \cite{SHT_inflation} it was shown that for MNL systems an interval of average network connectivities
$I=[k^-,\, k^+]$ exists so that $\lambda_1<0$ for
$\mk<k^-$ (Area A), $\lambda_1\sim0$ for $\mk\in I$ (Area B),  and $\lambda_1>0$ for $\mk>k^+$ (Area C).
The two values of $\mk$, which estimate the beginning and the end of the $\lambda_1(\mk)\sim 0$
{\it plateau}, i.e. the interval $I$, are 
given by $k^-=D^2$ and $k^+=2 D^2$.
It also was shown, that as $\mk$ gets larger than $D^2$, the number $N_{\rm zero}$ of nodes $i$, whose concentration $x_i=0$ 
increases monotonically, due to the positivity condition, until $N_{\rm zero}\sim N/2$. 
The sub-network of size $N_{\rm on}=N-N_{\rm zero}$, consisting of those nodes $j$ of the full network, which have nonzero concentration $x_j$, we call the {\it active} network. If two nodes, $i$ and $j$, of the active network have a link
$A_{ij}$ in the full network, then the active network inherits this {\it active} link. We denote the 
adjacency matrix of the active network with $A^{\rm on}_{ij}$.

Technically, the positivity condition is implemented 
so that $x_i(t+dt)=0$, if $x_i(t)+\dot x_i(t)dt \leq 0$.
In \cite{SHT_inflation} the positivity condition, 
Eq. (\ref{poscond}), was implemented in a slightly different way, i.e.
$x_i(t+dt)=x_i(t)$, if $x_i(t)+\dot x_i(t)dt \leq 0$.
The different implementation has a small effect on the plateau formation.

\section{Results}\label{results}

We now present (i) the Lyapunov spectrum of the MNL model, (ii) the probability to find 
growing, decaying, or stable dynamics and the size and topological properties of the {\it active} catalytic network.
Further we present (iv) the probability of finding oscillating 
time series and their characteristic frequencies. In all following figures the theoretical plateau interval 
$[k^-,\, k^+]$ is marked (gray shading).

\subsection{The Lyapunov spectrum}

It is straight forward to compute the full Lyapunov spectrum $\lambda_p$ 
which allows to determine properties of attractors in greater detail.
In particular we computed the {\it Kaplan-York Dimension} $D_{KY}$ which gives an upper bound for the information dimension of the system and the {\it Kolmogorov-Sinai Entropy} $E_{KS}$, see e.g. \cite{EckmannRuelle}. 
While $D_{KY}$ gives an estimate of the dimension of the attractor, i.e. the phase-space volume-preserving subspace of the dynamics, $E_{KS}$ can be interpreted as a measure of the number of excited states in the system. 

In Fig. (\ref{fig:LyapSpectrum}) we summarize numerical results. 
(a) shows the first ten Lyapunov exponents $\lambda_p$, $p=1\dots 10$, as functions of $\mk$.
Clearly, $\lambda_1\sim 0$ matches the theoretical plateau region, $[k^-,\, k^+]$.
(b) shows how the full Lyapunov spectrum depends on $\mk$. 
In area A ($\mk<k^-$) all $\lambda_p$ are densely arranged.
In area B (the plateau) $\lambda_1\sim 0$, while all $\lambda_p$ decrease monotonically 
for $p>1$. As a consequence, the Lyapunov spectrum gets less dense with growing $\mk$ and covers an increasing range
of negative values. In area C ($\mk>k^+$) the Lyapunov spectrum gets still less dense.
Yet, this decrease in density is qualitatively different than in area C. $\lambda_1$ is increasing in area C,
while $\lambda_p$, for large $p$, still decreases monotonically, but less pronounced than in area B. 
This can be seen clearly in Fig. (\ref{fig:LyapSpectrum}) (c-e), where Lyapunov spectra
for various values of $\mk$ are shown as functions of $p$.
In (c) the values of $\mk$ are chosen from area A, in (d) from area B, and 
in (e) from area C.
%

Further, Fig. (\ref{fig:LyapSpectrum}) shows the average 
Kolmogorov-Sinai entropy (f) and the average Kaplan-York dimension (g). 
Both quantities require the existence of some $\lambda_p>0$, i.e. $E_{KS}$ and $D_{KY}$ cannot be computed 
for area A. In area B and C averages of $E_{KS}$ and $D_{KY}$ can only be taken over 
realizations, which have at least some $\lambda_p>0$. 
Clearly, in area C the entropy per node $E_{KS}/N$ and the fraction $D_{KY}/N$ of 
the volume-preserving subspace seem to become independent of system-size $N$, for sufficiently large $\mk>k^+$.
However, in the plateau region, area B, both quantities do not seem to
scale with system size and finite size effects may
become relevant. It is an interesting open question towards which limit-function 
$E_{KS}/N$ and $D_{KY}/N$ converge as $N\to\infty$.

\subsection{Stability of MNL systems}

\begin{figure}[htbp]%
\hspace{1.5cm}\includegraphics[width=.8\columnwidth]{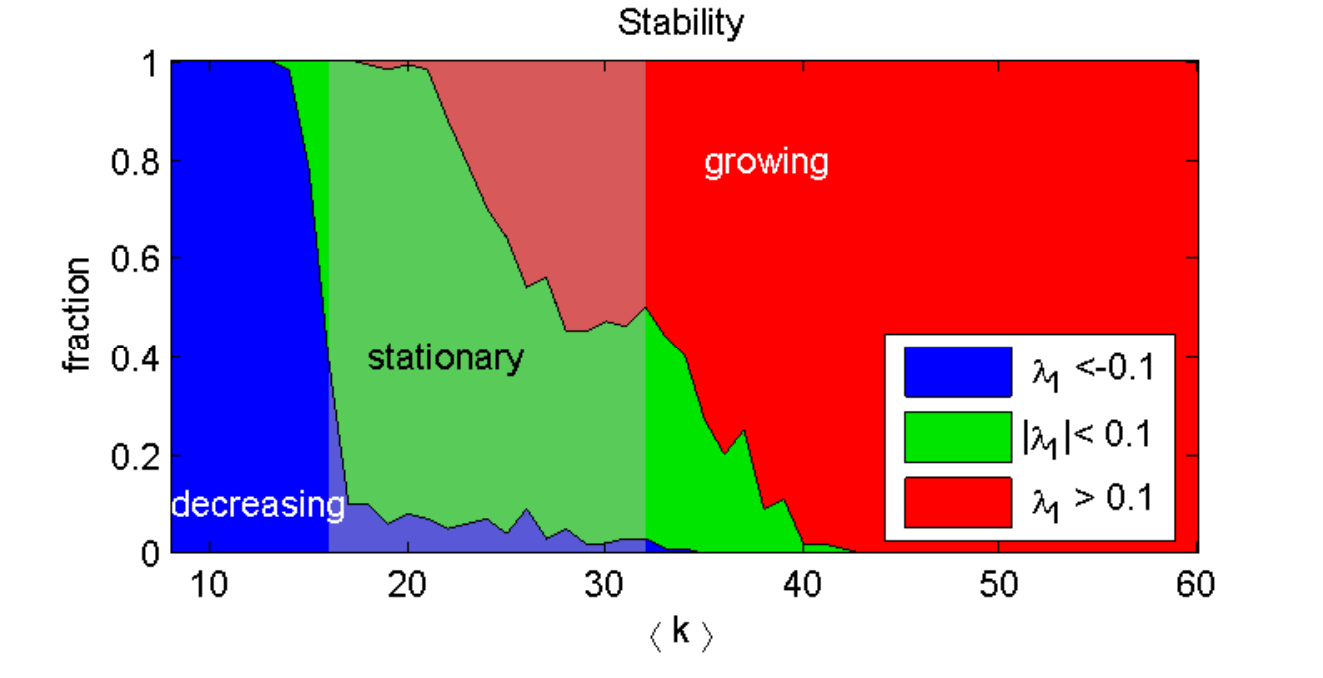}%
\caption{Fraction of realizations which lead to exponentially growing ($\lambda_1>0.1$), decaying ($\lambda_1<-0.1$) and stable time series ($|\lambda_1|\leq 0.1$) computed from 100 realizations, $N=500$, $D=4$, time interval, 
$[200,\, 1000]$, $\sigma=\bar{\sigma}=0.1$, $\sigma_A=1$, and $x^0$=1000.}
\label{fig:Stability}%
\end{figure}

What is the probability of finding the dynamics of MNL systems to be characterized (i) by exponential growth,
(ii) exponential decay, or (iii) non-exponentially growing stationary or oscillatory dynamics? 
Living systems can be expected to
exist close to stationary or oscillatory states \cite{pross,pross2}. 
Sufficiently positive and sufficiently negative MLEs, $\lambda_1$,
in complete analogy to linear systems, indicate 
exponential growth or decay. Therefore the probabilities of finding MNL systems in one of the 
growth modes (i-iii) can simply be estimated by thresholding $\lambda_1$, for a sufficiently small threshold 
$\ep>0$. Counting realizations in the MNL ensemble with (i) $\lambda_1\geq\ep$,
(ii) $-\ep\geq\lambda_1$, and  (iii) $\ep>|\lambda_1|$ estimates the ratios of growth mode fractions (i-iii).
Numerical results for $\ep=0.1$ are given in Fig. (\ref{fig:Stability}). 
Clearly, dynamics of type (iii) is favored in the plateau region. Yet, to a much lesser 
extent, stationary and oscillatory dynamics also can be found in areas A and C.

\subsection{The active genetic regulatory sub-network}

%
%
\begin{figure}[ht]%
\hspace{1.5cm}\includegraphics[width=0.8\columnwidth]{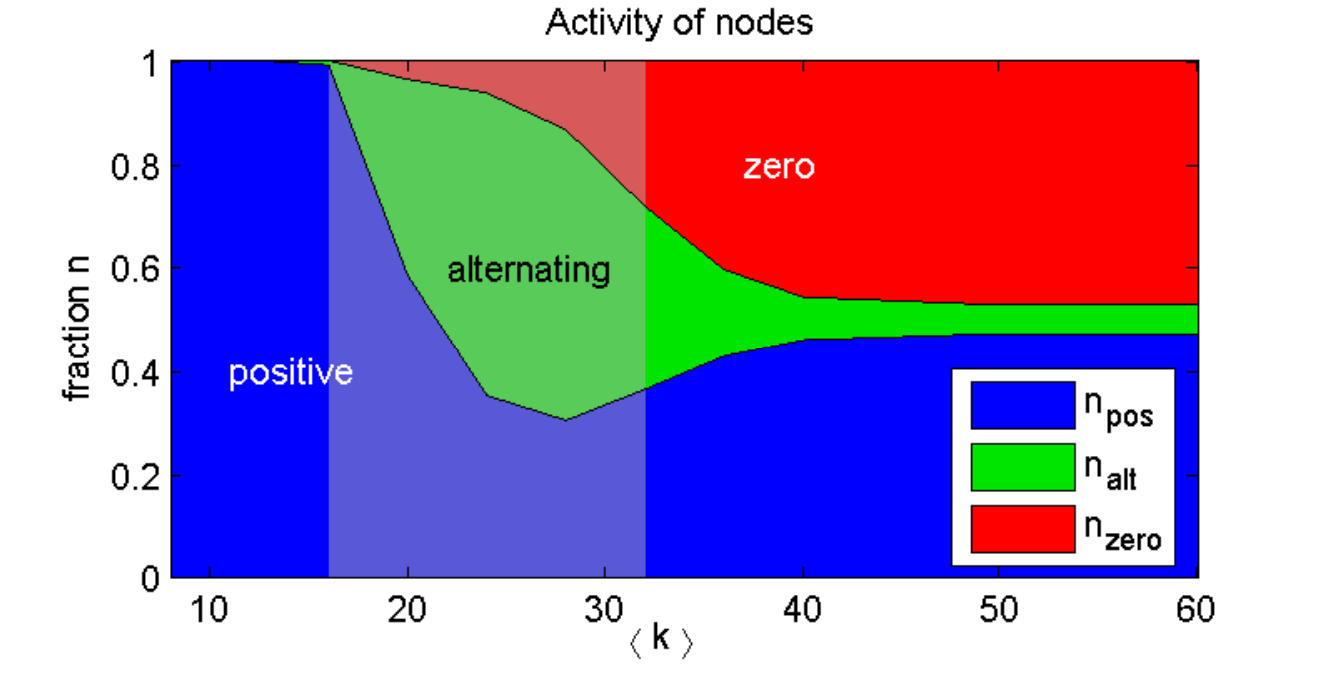}\\ \vspace{0cm}%
	\caption{   
Average fractions of $x_i$: positive ($n_{\rm pos}$), zero ($n_{\rm zero}$),
or alternating ($n_{\rm alt}$). Averages are taken over $1000$ realizations,
time interval, $[500,\, 1000]$, $N=500$, $D=4$, $\sigma=\bar{\sigma}=0$, $\sigma_A=1$, $x^0=1000$.
}\label{fig:Active_1}
\end{figure}
%
%
%
%
\begin{figure}[ht]%
\hspace{0.9cm}
\begin{minipage}[b]{0.9\textwidth}
\hspace{0.1cm}
\includegraphics[width=0.94\columnwidth]{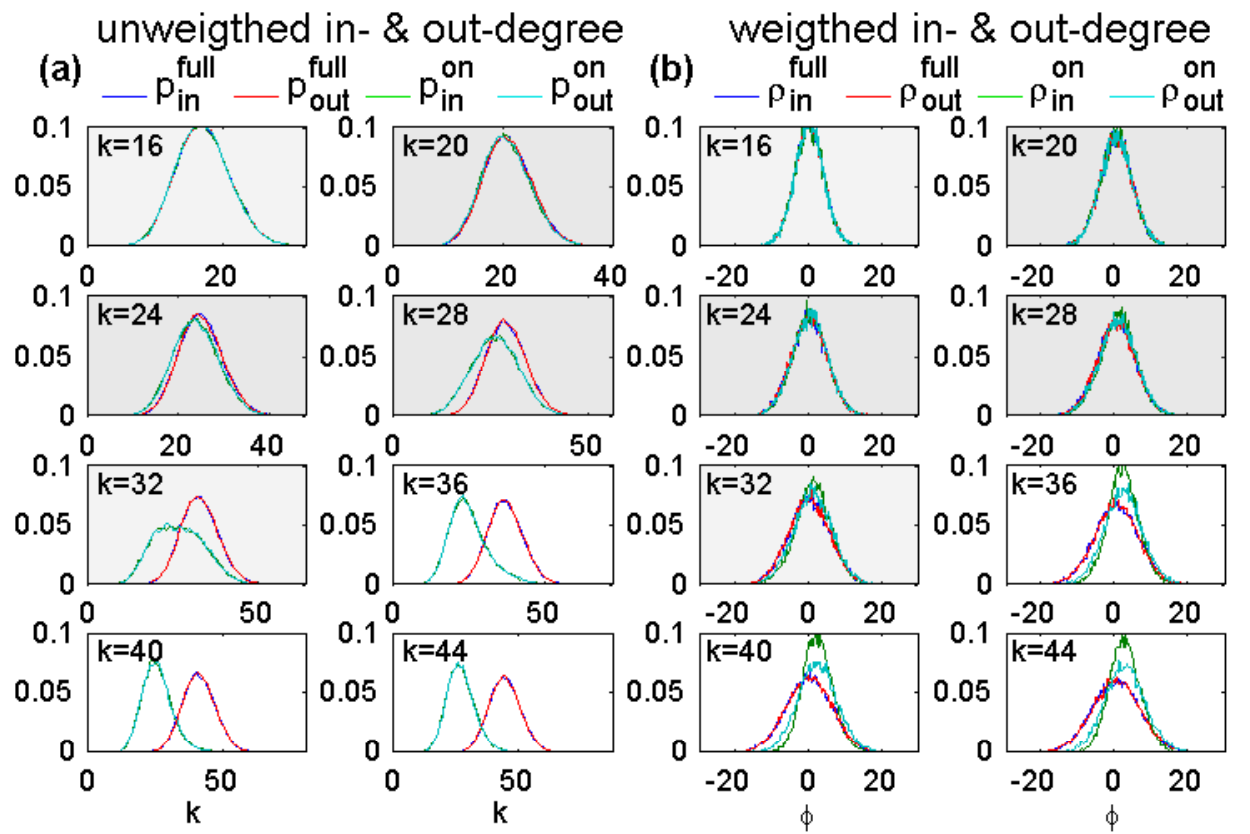} \\ \vspace{0.0cm}\hspace{-0.2cm}%
\includegraphics[width=\columnwidth]{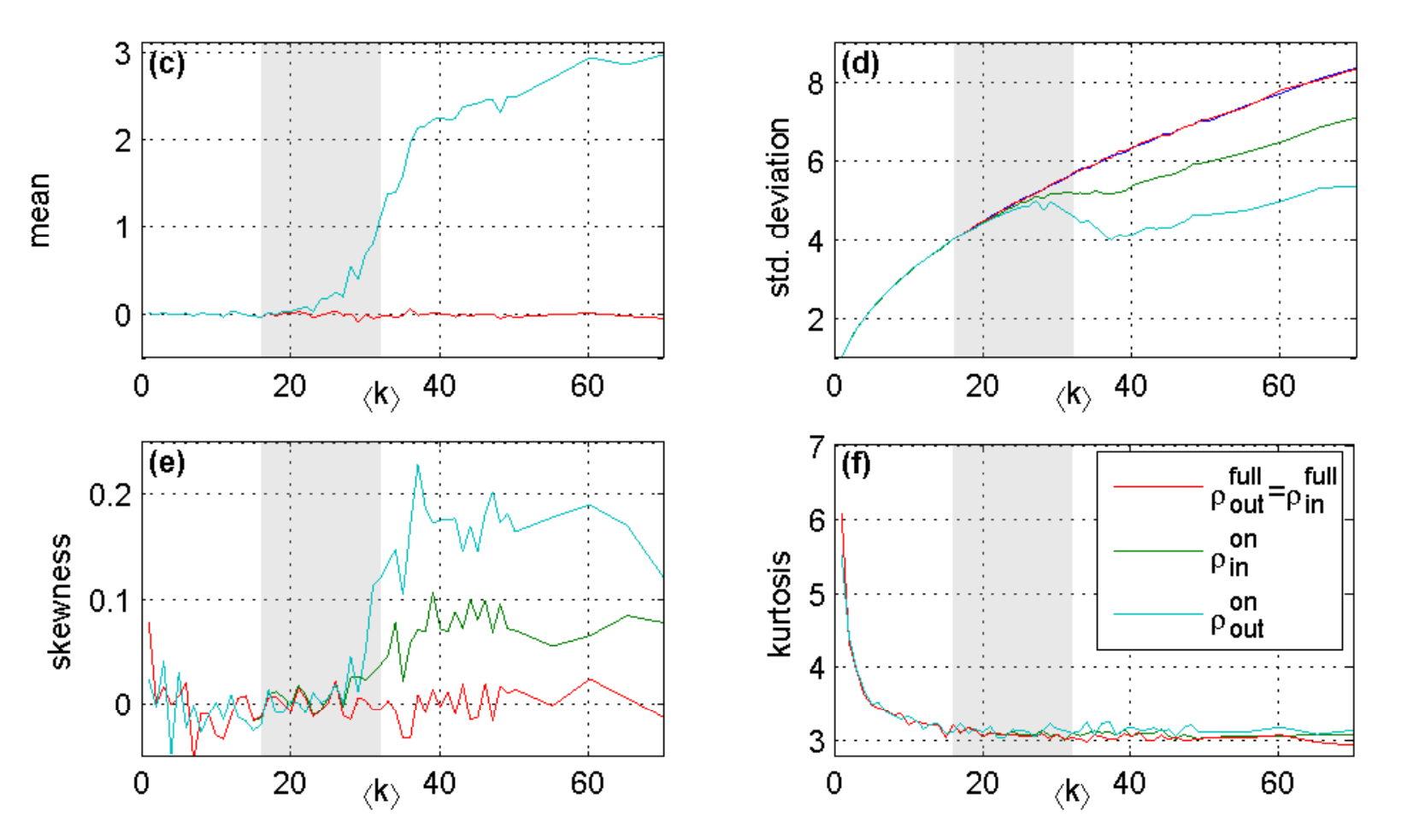} %
\end{minipage} 
	\caption{   
(a) Unweighted in- and out-degree distributions of the active regulatory sub-network for various $\mk$. 
Active in- and out-degree, $p^{\rm on}_{\rm in/out}(k)$, are practically indistinguishable. 
(b) Weighted in- and out-degree distributions. 
In- and out-weight distributions, $\rho^{\rm on}_{\rm in/out}(\phi)$, of active weights are clearly distinguishable.
$\phi=\sum A_{ij}$ and the sum runs over $i$ or $j$ for in- and out-weight distribution, respectively. 
(c) Mean, (d) standard deviation, (e) skewness and (f) kurtosis of the in/out-weight distributions. 
Differences between in- and out-weight distributions are found in the standard deviation and the skewness.
Averages are taken over 50 realizations, $N=500$, time interval, $[500,\, 1300]$, $D=4$, 
$\sigma=\bar{\sigma}=0.1$, $\sigma_A=1$, $x^0=1000$.  
}\label{fig:Active}
\end{figure}
%
%
Recent evidence from the analysis of genetic regulatory networks 
\cite{barabasi_xx, barabasi_x},
suggests topological differences between in- and out-degree distributions.
Can differences between
in- and out-degree distributions appear merely by the fact that
the full interaction network $A_{ij}$ is different from the {\it active} sub-network 
$A^{\rm on}_{ij}$? 
Is it possible that through a symmetry-breaking mechanism the active in- and out-degree distribution 
$p_{\rm in}^{\rm on}(k)$ and $p_{\rm out}^{\rm on}(k)$ become different from $p^{\rm full}(k)$?

We have analyzed the average properties of active sub-networks of the MNL model 
and distinguish three types of nodes in MNL systems: 
(i) nodes $i$ with concentrations $x_i>0$ for all times,
(ii) nodes $j$ with $x_j=0$ for all times, and
(iii) nodes that alternate between on and off.
The associated numbers of nodes are $N_{\rm pos}$, $N_{\rm zero}$, and $N_{\rm alt}$, 
where $N=N_{\rm pos}+N_{\rm zero}+N_{\rm alt}$ and 
the fraction of nodes are denoted $n_{\rm pos}=N_{\rm pos}/N$, $n_{\rm zero}=N_{\rm zero}/N$, 
and $n_{\rm alt}=N_{\rm alt}/N$.
Note that the size of the active sub-network is $N_{\rm on}=N_{\rm pos}+N_{\rm alt}$.
In Fig. (\ref{fig:Active_1}) we show these fractions for a network with $N=500$.
Alternating nodes are most important
in the plateau region, where $N_{\rm zero}$
starts to grow, i.e. the active networks shrinks with growing $\mk$.
For $\mk>k^+$ the fraction $n_{\rm alt}$ decreases and reaches a constant value $n_{\rm alt}\sim0.05$;   
$n_{\rm pos}$ and $n_{\rm zero}$ become equally large. 
 
In Fig. (\ref{fig:Active} (a) {\it unweighted in- \& out-degree}
shows the in- and out-degree distributions 
of the active network for various $\mk\geq k^-$. 
The degree distribution of the full network is shown (red) for reference.
Although in- and out-degree distribution of the active network differ substantially from the degree distribution
of the full network, in- and out-degree distributions essentially remain identical.
If we look at the weight distributions, $\rho^{\rm on}_{\rm in}$ and $\rho^{\rm on}_{\rm out}$,  
associated with active in- and out-links in (b) {\it weighted in- \& out-degree}
the situation changes: differences in the in- and out-weight distributions begin to show. 
These differences are recognizable in (d) the standard deviation and (e) the skewness
of the weight distributions, but not in (c) the mean and (f) the kurtosis
of the active weight distributions.
This establishes evidence that a possible symmetry breaking of in- and out-degree distributions of 
complete chemical reaction networks can arise due to the natural nonlinearity in the dynamics 
of the chemical reactive systems, i.e. $x_i\geq0$ for all $i$, at all times. 
However, the size of this effect seems to be insufficient to explain 
the size of topological differences 
\cite{barabasi_xx, barabasi_x}.

\subsection{Oscillating modes in MNL systems}

%
%
\begin{figure}[ht]%
\begin{tabular}{c c}
\begin{minipage}[b]{0.65\textwidth}
\includegraphics[width=\columnwidth]{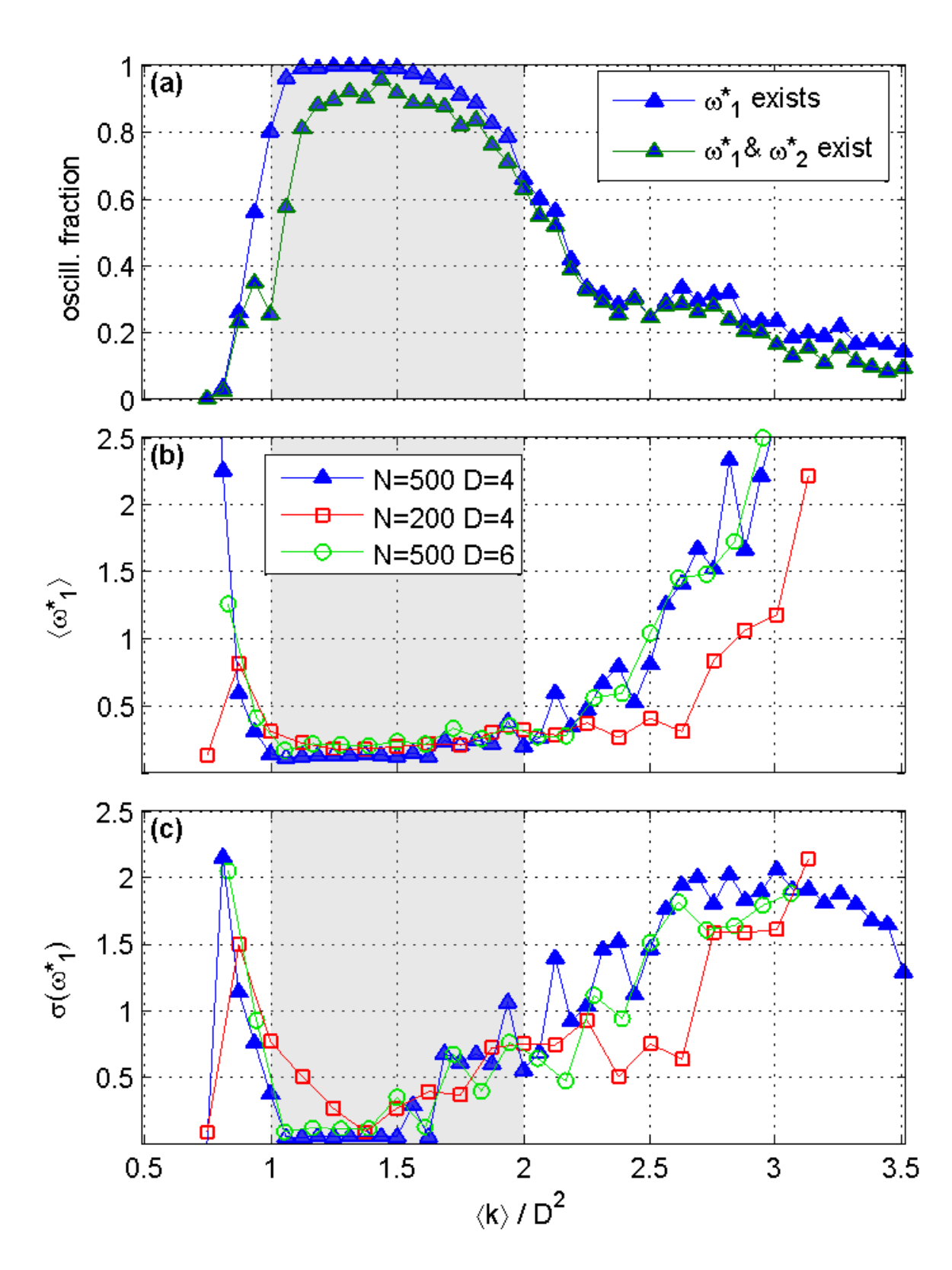} 
\end{minipage} & \hspace{0.1cm}
\begin{minipage}[b]{0.25\textwidth}
\caption{
(a) Probability of finding oscillating realizations: with existing fundamental frequency $\omega^*_1$ (blue)
and both, existing $\omega^*_1$ and $\omega^*_2$ (green). 
(b) Average $\omega^*_1$ as a function of $\mk$.
(c) Standard deviation of $\omega^*_1$. 
$N=500$, time interval, $[1000\,,\, 3000]$, 
$D=4$, $\sigma=\bar{\sigma}=0$, $\sigma_A=1$, $x^0=1000$.
In (b) and (c) 
$N=500$, $D=6$, (green circles) and
$N=200$, $D=4$ (red squares) are shown for comparison. 
\vspace*{4.5cm}
}
\label{fig:Frq}
\end{minipage}
\end{tabular}	
\end{figure}
%
%

What is the fraction of MNL systems, which display oscillating dynamics and what
are their typical frequency distributions? 
We find that if a particular realization of an MNL system shows oscillating dynamics, then all 
$x_i$ in the active network of the particular realization follow the same fundamental oscillation-pattern. 
The dominant frequencies $\omega^*_s$, $s=1,2,\dots s_{max}$,
correspond to local maxima in the power-spectra of the active $x_i$. 
$s_{max}\leq N$ is the maximal number of detectable local maxima in the power-spectrum of the
MNL system dynamics. 
We looked for fundamental frequencies $\omega^*_1$ (if existing).

Technically we identified $\omega^*_1$ and $\omega^*_2$ in the following way. 
We computed the power-spectrum $P_i(\omega)$ 
for each time series $x_i(t)$ in a particular realization.
We took the weighted average $g(\omega)\equiv\langle w_i P_i(\omega) \rangle_{i}$
over all nodes of the realization. 
The weights, $w_i=P_i(\omega_1)^{-1}$, have been chosen inverse-proportional to 
the power $P_i(\omega_1)$ of the lowest frequency $\omega_1$ in the power-spectrum.
This choice turned out to be optimal for correctly detecting the dominant frequencies
$\omega^*_s$ of MNL dynamics. 
The first frequency $\omega^*_1$ can be found in the following way.
We have searched for the local minimum of $g(\omega)$ with the smallest frequency
$\hat\omega_1$. If no such local minimum exists the time series was classified as non-oscillating.
If $\hat\omega_1$ exists, the fundamental frequency, $\omega^*_1>\hat\omega_1$, is determined such that 
$g(\omega^*_1)$ is the maximum of all $g(\omega)$, with $\omega>\hat\omega_1$.   
Similarly, we computed a second dominant frequency by searching for the next local minimum 
$\hat\omega_2>\omega^*_1$, and take $g(\omega^*_2)$ to be the maximum of all 
$g(\omega)$, with $\omega>\hat\omega_2$. If $\hat\omega_2$ exists, $\omega^*_2$ is the second
characteristic frequency of the system. 

In Fig. (\ref{fig:Frq}) (a) shows the probability of finding a realization of the MNL model  
possessing a fundamental frequency and a second dominant frequency. 
Oscillating realizations are dominant in the plateau region and the probability of finding 
oscillating realizations is close to certainty for $k^-<\mk<(k^-+k^+)/2$.
For $\mk>(k^-+k^+)/2$ this high probability decreases, but still has a value of about $0.6$
for $\mk=k^+$. 
Furthermore, in Fig. (\ref{fig:Frq}) (b) the average and (c) the standard-deviation
of the fundamental frequencies $\omega^*_1$ are shown.  


\section{Conclusions}\label{conclusions}

We presented results on properties of the MNL model.
We analyzed the Lyapunov spectrum of the model and computed the 
Kolmogorov-Sinai Entropy and the Kaplan-York Dimension, characterizing the attractors 
of MNL dynamics.
We analyzed stability properties by computing the probabilities for finding 
exponentially growing, decaying and non-exponentially growing (stable) dynamics and found that
stable dynamics plays a dominant role in the plateau interval, $[k^-,\, k^+]$. 
We determined characteristic 
fractions of concentration levels, which are always down-regulated to zero, are always positive,
or are alternating, i.e. oscillating between zero and non-zero concentration levels. 
Nodes with alternating concentration levels are dominating in the plateau interval.   
We analyzed topological properties of the active regulatory network,
consisting only of molecular species (nodes) with nonzero concentration levels in a given time period. 
We found no symmetry-breaking in the in- and out-degree distributions of the active regulatory network
with respect to the full network $A_{ij}$. 
However, we found symmetry-breaking in the in- and out-weight distributions of active networks. 
This indicates that in chemically reactive systems the natural nonlinearity introduced by 
the positivity condition, i.e. concentrations of molecular species can
never be negative, suffices to implement a symmetry-breaking in the topology 
of the system, which can actually be measured. One may speculate if the pronounced differences of in- and out-degree distributions as found in 
living organisms, have their origin in symmetry-breaking mechanisms, which later could become amplified by 
selective evolutionary processes.
Finally, we determined probabilities of finding oscillating dynamics
in MNL systems and analyzed fundamental properties of their dominant frequencies.
We found that oscillatory dynamics is most likely, in fact almost certain, 
for average connectivities of networks chosen from the plateau interval. 
This corresponds well to the observation that regulatory networks of living organisms, cells in particular, 
frequently show sub-networks with oscillatory dynamics. 
The properties analyzed indicate that near the {\it edge of chaos}
MNL system -- despite the simplicity of the MNL model -- display 
many important characteristic properties, which are expected from living matter.

\ack Supported by Austrian Science Fund FWF project P19132. 


\end{document}